# Temperature and Frequency Dependence of Complex Conductance of Ultrathin YBa$_2$Cu$_3$O$_{7-x}$ Films: A Study of Vortex-Antivortex Pair Unbinding

V.A. Gasparov[*], G. Tsydynzhapov[*], I.E. Batov[**] and Qi Li[***]

*Institute of Solid State Physics RAS, 142432, Chernogolovka, Moscow distict., Russian Federation
** University of Erlangen-Nürnberg, Erwin-Rommel-Str.1, D-91058 Erlangen, Germany
***Department of Physics, Pennsylvania State University, University Park, PA 6802, USA

*We have studied the temperature dependencies of the complex sheet conductance, $\sigma(\omega,T)$, of 1-3 unit cell (UC) thick YBa$_2$Cu$_3$O$_{7-x}$ films sandwiched between semiconducting Pr$_{0.6}$Y$_{0.4}$Ba$_2$Cu$_3$O$_{7-x}$ layers at high frequencies. Experiments have been carried out in a frequency range between: 2 - 30 MHz with one-spiral coil technique, 100 MHz - 1 GHz frequency range with a new technique using the spiral coil cavity and at 30 GHz by aid of a resonant cavity technique. The real, ReM(T), and imaginary parts of the mutual-inductance $M(T,\omega)$, between a coil and a film, were measured and converted to complex conductivity by aid of the inversion procedure. We have found a quadratic temperature dependence of the kinetic inductance, $L_k^{-1}(T)$, at low temperatures independent of frequency, with a break in slope at $T^{dc}_{BKT}$, the maximum of real part of conductance, $\omega\sigma_1(T)$, and a large shift of the break temperature and the maximum position to higher temperatures with increasing frequency $\omega$. We obtain from these data the universal ratio $T^{dc}_{BKT}/L_k^{-1}(T^{dc}_{BKT})$ = 25, 25, and 17 nHK for 1-, 2- and 3UC films, respectively in close agreement with theoretical prediction of 12 nHK for vortex-antivortex unbinding transition. The activated temperature dependence of the vortex diffusion constant was observed and discussed in the framework of vortex-antivortex pair pinning.*

## 1. INTRODUCTION

Although, many observations of the Berezinski-Kosterlitz-Thouless (BKT) transition in YBCO[1-12], BiSrCaCuO[13,14], and TlBaCaCuO[15] compounds have been reported, detailed comparison of the experimental data with the theory by Davis *et al.*[16] showed disagreements possibly due to inhomogeneity and vortex pinning. Rogers *et al.* reported that the usual BKT transition, i.e. all thermally activated vortices form vortex-antivortex pairs at temperatures below BKT transition temperature T$_{BKT}$, was not observed in ultrathin Bi$_2$Sr$_2$Cu$_2$O$_8$ films from a low-frequency noise measurement due to vortex pinning[17]. Repaci et al.[18] showed from the study of *dc* I-V curves that free vortices exist even at low temperatures, indicating the absence of the BKT transition. They have pointed out that a precondition[19] for the BKT transition to occur in a superconductor, i.e. the sample size $L_s<\lambda_{eff}$, where $\lambda_{eff}=2\lambda^2/d$ is the effective penetration depth and d is the film thickness, is not satisfied in YBCO films as thin as one unit cell.

According to the BKT theory extended to finite frequencies[20-22], higher frequency currents sense vortex-antivortex pairs of smaller separations. At high frequency, the electromagnetic response of a 2D superconductor is dominated by those bound pairs that have r ~ $l_\omega$, where $l_\omega = (14D/\omega)^{1/2}$ is the vortex diffusion length and D is the vortex diffusion constant. Using the Bardeen-Stephen formula for free vortices, $D = 2e^2\xi_{GL}^2 k_B T/\pi\hbar^2\sigma_n d$ [23], we estimate that $l_\omega < 1$ μm at ω>10 MHz, which is much less than $\lambda_{eff}$ ~ 40 μm for the 1-UC YBCO film. This implies that it is possible to detect the response of vortex-antivortex pairs with short separation lengths at high frequencies in the samples even though the usual BKT transition is not present as shown in *dc* and low frequency measurements.

In this paper, we report the frequency and temperature dependences of the complex sheet conductance, $\sigma(\omega,T)$, of 1-UC to 3-UC thick YBCO films sandwiched between semiconducting $Pr_{0.6}Y_{0.4}Ba_2Cu_3O_{7-x}$ layers in a frequency range between 1 MHz to 30 GHz. Here $\sigma(\omega,T) = \sigma_1(\omega,T) - i[\omega L_k(\omega,T)]^{-1}$, where $\sigma_1(\omega,T)$ is the dissipative component of the sheet conductance and $L_k(\omega,T) = \mu_0\lambda^2/d$ is the sheet kinetic inductance ($\mu_0$ - the permeability of free space). In the frequency range of our measurements, the condition $l_\omega \sim r < \lambda_{eff}$ is met. Preliminary results of this study have been published in Ref. 24.

We found a large increase of $T_{BKT}(\omega)$ as a function of frequency for those films from 4 MHz to 30 GHz. A jump of $L_k^{-1}(T)$, a maximum of $\sigma_1(T)$ at $T^\omega_{BKT}$, and the scaling of the universal superfluid jump close to the theoretical prediction at the frequencies studied was observed. The superfluid jump is suppressed in a very small magnetic field. Vortex pinning with thermally activated vortex diffusion constant was found in the samples at low frequencies, while it did not destroy the vortex-antivortex pairs with short separation lengths, i.e. at microwaves.

## 2. EXPERIMENTAL SETUP

Ultrathin YBCO layers sandwiched between 100 Å buffer and 50 Å cover layers of $Pr_{0.6}Y_{0.4}Ba_2Cu_3O_{7-x}$ were grown epitaxially on (100) $LaAlO_3$ substrates using a multitarget pulsed-laser deposition system. The samples were made at identical conditions as the sample used in Ref.18 for *dc* measurements and the detailed information can be found elsewhere[24-26]. The three films thickness we examined were nominally 1-, 2-, and 3- unit cells thick and had the *c* axis normal to the film surface. The samples were made at different oxygen composition and therefore we will call them as S1 and S2 ones. The contacts were made at the edges of the 1×1 $cm^2$ 3UC S1 film for van der Pauw four-point resistance measurements.

The $\sigma(\omega,T)$ at RF in thin films was investigated employing a single coil mutual inductance technique. This technique, originally proposed in[27] and lately improved in [28], has the advantages of the well known two-coil geometry[29,30], and was extensively used for the study of the λ(T) dependence for YBCO and $MgB_2$ films[9,28,31,32]. In this radio frequency technique, the change of inductance ΔL of a one-layer pancake coil located in the proximity of the film and connected in parallel with a capacitor C is measured. The LC circuit is driven by the impedance meter (VM-508 TESLA) operating at 2-30 MHz, with a high frequency stability of 10 Hz. The film is placed at small distance (~0.1 mm) above the coil in vacuum, which allows the sample temperature to vary from 4.2 K to 100 K, while the LC circuit is kept at 4.2 K. This eliminates any contribution from changes in L and C as function of temperature from interfering with the measurements. The complex mutual inductance M between the coil and the film can be obtained through:

$$\mathrm{Re}\, M(T) = L_0 \cdot (\frac{f_0^2}{f^2(T)} - 1) \qquad (1)$$

$$\mathrm{Im}\, M(T) = \frac{1}{2\pi f(T)C^2} \cdot [\frac{1}{Z(T)} - \frac{1}{Z_0(T)} \cdot \frac{f^2(T)}{f_0^2}] \qquad (2)$$

Here L, Z(T), f(T), $L_0$, $Z_0$ and $f_0$ are the inductance, impedance and the resonant frequency of the circuit with and without the sample, respectively. In the low frequency regime, where the coil wire diameter is much thinner than the skin depth at the working frequency, the expression of the variation of the M(T) (relative to the case where no sample is in the coil, $M_0$, as a function of the $\sigma(T)$ may written as:

$$\Delta M(T) = \pi \mu_0 \cdot \int_0^\infty \frac{M(q)}{1 + 2ql \cdot \coth(\frac{d}{l})} dq \, , \qquad (3)$$

where M(q) plays the role of mutual inductance at a given wave number q in the film plane and depends on the sample-coil distance h, d is the sample thickness, and l is a complex length defined as l= $[1/(i\omega\mu_0\sigma_1+\lambda^2)]^{1/2}$, (more details can be found in [28]). A change in real, ReM(T), and imaginary, ImM(T), parts of M(T) were detected as a change of resonant frequency f(T) of the oscillating signal and impedance Z(T) of the LC circuit, and converted into $L_k(T)$ and $\sigma_1(T)$ by both, using Eqs.1, 2 with inversion mathematical procedure and Eq.3. The MW measurements were performed using the cavity formed with a similar spiral coil with no capacity in parallel. The coil form the radio frequency resonator coupled to a two coupling loops and is driven by the radio-frequency signal generator/receiver (from 100 MHz to 1 GHz). In this case the quality factor of the resonator Q and the resonance frequency were measured and converted to ReM(T) by Eq.1 and to ImM(T) by:

$$\mathrm{Im}\, M(T) = L_0 \cdot \frac{f_0}{f(T)} \cdot [\frac{1}{Q(T)} \cdot \frac{f_0}{f(T)} - \frac{1}{Q_0}] \qquad (4),$$

where Q (T) and $Q_0$ (T) are the quality factors with and without the samples, respectively.

The MW losses were measured using a resonant cavity technique with the gold-plated copper cylindrical cavity operated in the TE011 mode at 29.9 GHz. The samples were mounted as a part of the bottom of the cavity through a thin gold-plated Cu-diaphragm with a small (3.5 mm) central hole. We assume the MW electric field E in the film to be uniform along normal direction and equal to the E without the film. The $\sigma_1(\omega,T)$ is thus proportional to $\Delta Q^{-1}(T) = Q^{-1}(T) - Q_0^{-1}(T)$ [24], where Q(T) and $Q_0(T)$ are the quality factors with and without the samples, respectively. The experimental scheme for measuring MW losses was based on the amplitude technique [24].

### 3. RESULTS AND DISCUSSION

Figure 1 displays the ReM(T) and ImM(T) curves for a 2UC S2 film at three different frequencies from 3 MHz to 500 MHz as measured by different techniques: LC circuit and spiral coil resonator. The MW data are normalized to RF

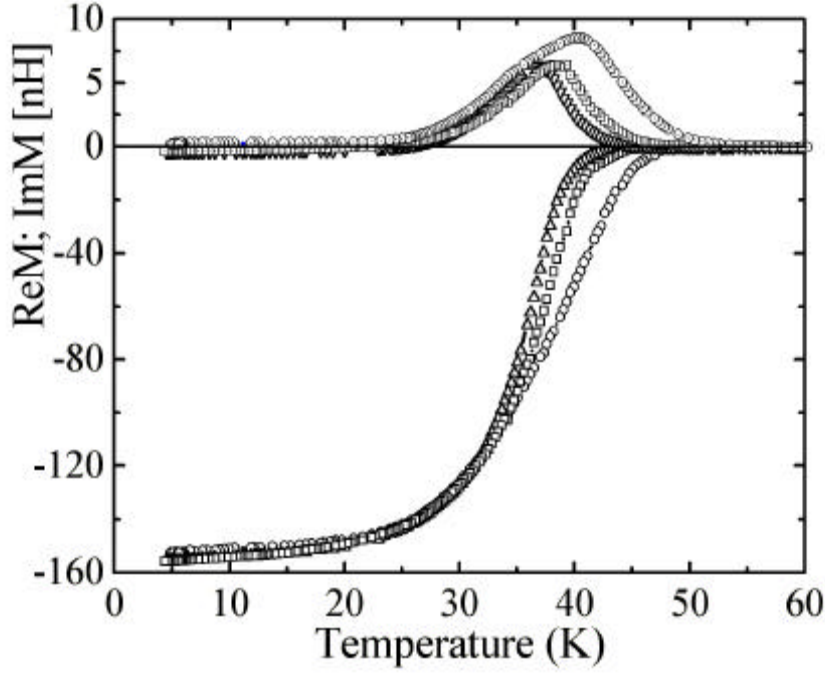

Fig.1. The ReM(T) and ImM(T) curves for a 2-UC film (S2) at different frequencies: 3 MHz (triangles), 26 MHz (squares) and 500 MHz (circles) calculated from raw f(T), Z(T) and Q(T) data. The solid lines describe a guide for the eye.

ImM(T) for 3UC S2 film measured at the same frequencies are shown in Fig. 2 as a function of temperature. The most noticeable feature of these data is rather high shift of the onset point $T_{c0}$ of ReM(T) transition with frequency, not observed in such measurements on thick films. Notice also, that the inductive response, ReM(T), starts at lower temperatures than ImM(T), characterized by a peak close to transition, and this shift is raised with frequency. We have carried out the mutual inductance measurements on $Pr_{0.6}Y_{0.4}Ba_2Cu_3O_{7-x}$ films and observed no any features in the temperature dependences of the mutual inductance M(T).

The ReM(T) and ImM(T) data are converted to $L_k^{-1}(T)$ and $Re\sigma(T)$ using Eq.3 and the mathematical inversion procedure[24] based on the same approach as in the two-coil mutual inductance method. Fig.3 shows the $L_k^{-1}(T)$ curves in very low perpendicular magnetic fields, and zero field $\omega Re\sigma(T)$ for the 1-UC and 2-UC films (S1). We found that $L_k^{-1}(T)$ fit well over a wide temperature range by a parabolic dependence[33]: $L_k^{-1}(T) = L_k^{-1}(0)[1-(T/T_{c0})^2]$, shown as thin solid lines in Fig.3. We emphasize that this quadratics equation fit the data below characteristic temperature which we define as $T^{\omega}_{BKT}$, and which coincide with the positions of the peaks in $\omega Re\sigma(T)$. The mean field transition temperature, $T_{co}$, determined by extrapolation of $L_k^{-1}(T)$ to 0, is larger than the onset of transitions of $L_k^{-1}(T)$, while is close to the onset point of $\omega Re\sigma(T)$ curves. Also, the $L_k^{-1}(0)$ fitted data are the same for H=0 and 5 mT while have different $T_{c0}$.

In Fig.4, we plot $\omega Re\sigma(T)$ at 8 MHz and $\Delta Q^{-1}(T)/Q_0^{-1}(4.2K)$ determined from MW data (30 GHz) for 3-UC sample (S1). The *dc* resistive transition of the same sample is also shown in the figure. According to the Coulomb gas scaling model, the resistance ratio $R/R_n$ is proportional to the number of free vortices and should follow a universal function of an effective temperature scaling variable $X = T(T_{c0}-T^{dc}_{BKT})/T^{dc}_{BKT}(T_{c0}-T)$ [20], which can be approximated by $\rho/\rho_n = C_0 X \exp[-C_1(X-1)^{-1/2}]$ (here $\rho_n$ is the normal state resistivity, $C_0$=1.7 and $C_1$=4.9 are constants). We plot T/X(T) as a function of T obtained by fitting $\rho/\rho_n$ data for the 3-UC film with this equation in the insert of Fig. 4. The best fit was found with $T_{c0}=78.5$ K (the point where $T/X=0$) and $T^{dc}_{BKT}= 56$ K. Here $T^{dc}_{BKT}$ represents a nominal *dc* BKT transition temperature which is lower then $T^{\omega}_{BKT}$ determined from RF and MW data.

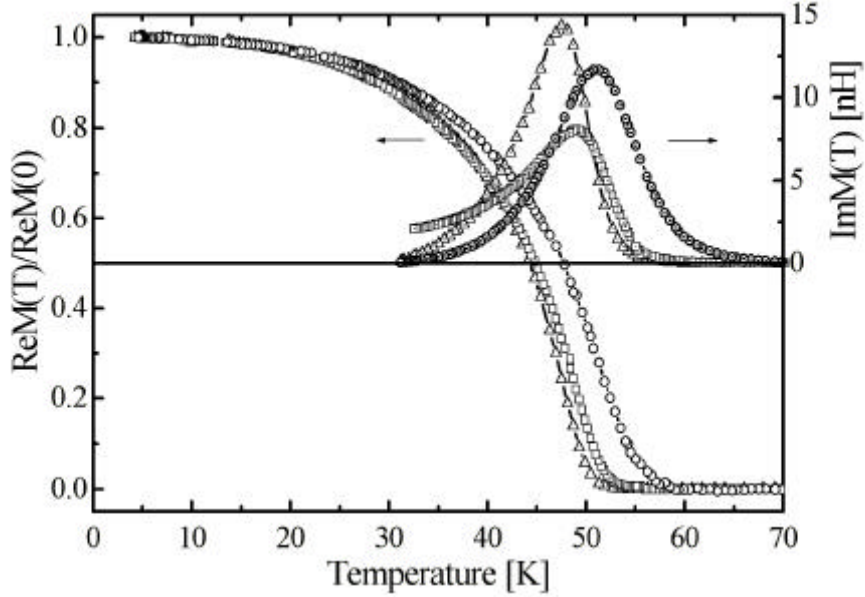

Fig.2. The ReM(T)/ReM(0) and ImM(T) curves for a 3-UC film (S2) at different frequencies: 3 MHz (triangles), 25 MHz (squares) and 500 MHz (circles) calculated from raw f(T), Z(T) and Q(T) data. The solid lines describe a guide for the eye.

There are three major features observed in our RF and MW measurements: (i) the large frequency dependence of $T_c(\omega)$, (ii) a foot jump in temperature dependence of the $L_k^{-1}(T)$, which is destroyed in weak magnetic fields, and (iii) a maximum in $\omega Re\sigma(T)$ with the onset of transition at higher temperatures than that of the $L_k^{-1}(T)$.

Indeed, as we can see from Fig.1, the $T^{\omega}_{BKT}$ value, determined as the maximum position of losses, increases on 4 K (from 36.6 K to 40.7 K) as the frequency raise from 3 to 500 MHz for 2-UC YBCO S2 film. The $T_{c0}$ value shift at 30 GHz is much larger: 74 K as compared with 61.5 K at 8 MHz (see Fig. 4). Even larger shift of $T^{\omega}_{BKT}$ was observed in 1-UC film and no shift was detected in a 2000 Å thick YBCO film. Although dissipation peaks have been observed in other systems when the skin depth is in the order of the sample size, the estimated skin depth of our samples is several orders of magnitude smaller than the sample size, indicating that the observed loss peak is not due to the skin effect[34].

The qualitative explanation of the frequency dependence of $T_{BKT}(\omega)$ and a peak in $\sigma_1(T)$ is as following. By probing the system at finite frequencies, the observed bound-pair response is dominated by those pairs with $r\sim l_\omega$. At temperatures below $T^{dc}_{BKT}$, the dissipation is proportional to the number of such vortex-antivortex pairs[20]. This number grows gradually with temperature up to $T^{\omega}_{BKT}$. At the higher temperature side, $\sigma_1$ decreases with increasing temperature since $\sigma_1\sim 1/(n_f\mu)$, where $n_f$ is the density of free vortices and $\mu$ is the vortex mobility[20]. Dissipation is largest when the correlation length $\xi_+(T)$, i.e. the average distance between thermally induced free vortices above $T^{dc}_{BKT}$, becomes equal to $l_\omega(T)$ which determines the BKT transition temperature at a given frequency $T^{\omega}_{BKT}$.

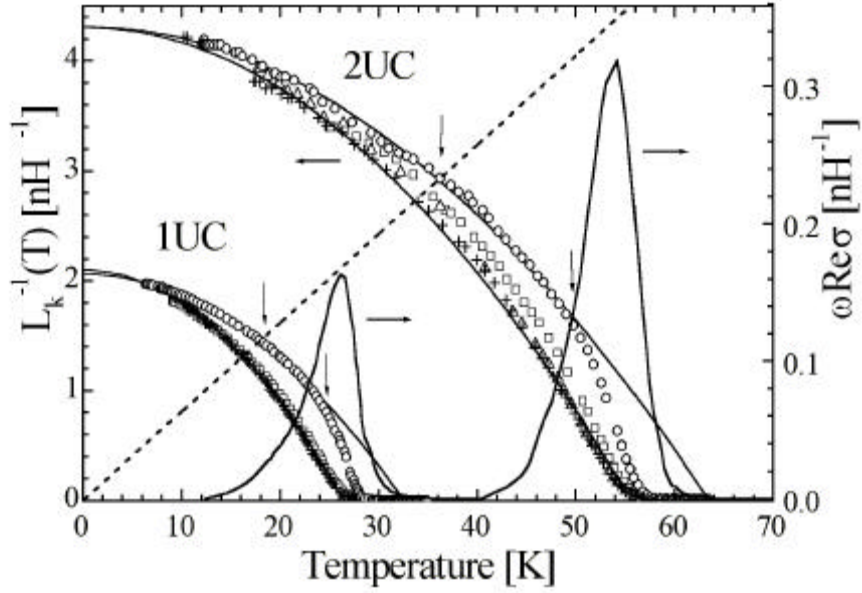

Fig.3. Temperature dependence of $L_k^{-1}(T)$ for 1- and 2-UC films (S1) at 8 MHz and different magnetic fields: 0 mT (circles), 2 mT (squares), 3 mT (triangles) and 4 mT (crosses). The solid lines curves shows $\omega \text{Re}\sigma(T)$ at zero field. The thin solid lines are quadratic fits to $L_k^{-1}(T)$ below $T^{dc}_{BKT}$ and for magnetic field data. Also shown is the theoretical BKT function (dashed line).

In order to see whether this assumption is correct, we plot theoretical BKT function $L_k^{-1}(T)$ as dashed straight line on Fig. 3 derived from the universal relationship:

$$K_R = \frac{d}{\lambda^2} \frac{\hbar^2 c^2}{16\pi e^2 k_B T_{BKT}} = \frac{2}{\pi} \qquad (5)$$

predicted by theory[4,16]. Notice however, that this theoretical dependence is valid for

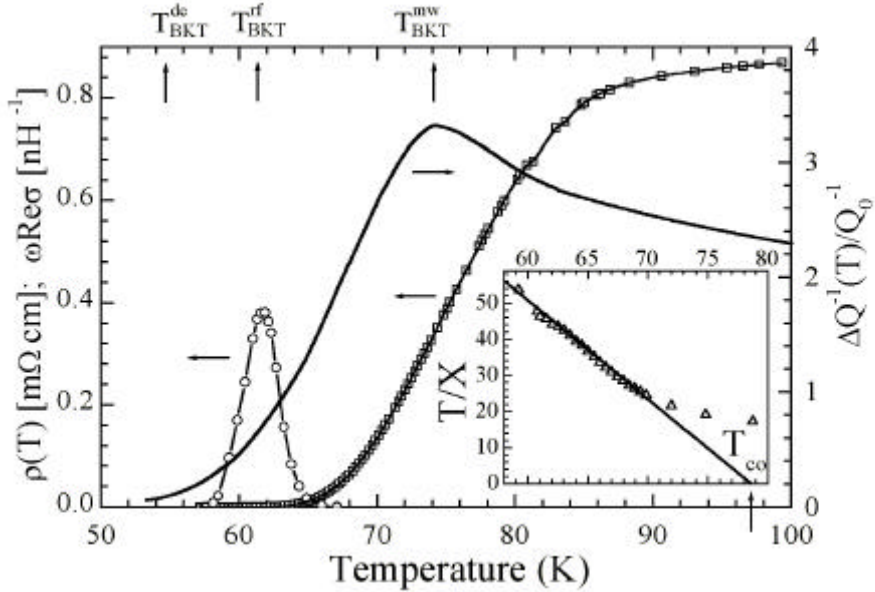

Fig.4. Temperature dependence of *dc* $\rho(T)$, $\omega \text{Re}\sigma(T)$ at 8 MHz and $\Delta Q^{-1}(T)/Q_o^{-1}$(4.2 K) at 30 GHz for a 3-UC sample (S1). Inset shows the universal plot: T/X vs T. Arrows indicate the $T_{BKT}$ values determined from the *dc* resistivity, loss function at 8 MHz and MW measurements (30 GHz).

*dc* case, while the frequency dependence of $T^{\omega}_{BKT}$ is obvious from the picture discussed above. This is why the critical temperatures determined from the intercept of dashed theoretical line with experimental $L_k^{-1}(T)$ is lower then the peak position of $\sigma_1(T)$. To see whether this description is correct, we plot the dependence of the penetration depth $\lambda^{-2}(T)$ derived from $L_k^{-1}(T)$, versus scaling variable - normalized temperature $T/T^{\omega}_{BKT}$ on Fig. 5. It is obvious, that all data for three studied S1 films at 8 MHz fall on the same curve, which proof of our definition of $T^{\omega}_{BKT}$ as the peak position of $\sigma_1(T)$.

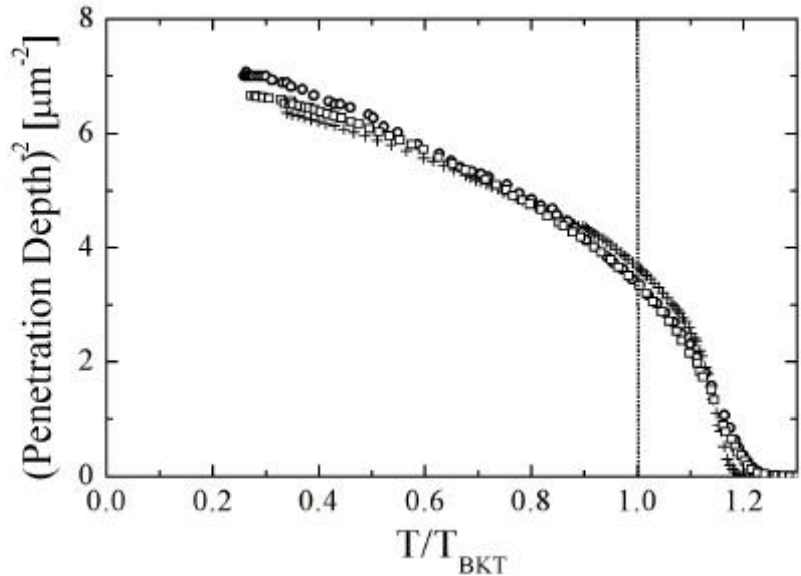

Fig.5. The penetration depth $\lambda^{-2}(T)$ derived from $L_k(T)$ vs normalized temperature $T/T^{\omega}_{BKT}$ at 8 MHz for: 1UC (circles), 2UC (squares) and 3UC (crosses) S1 films.

A central quantity in the dynamic description of BKT transition is the frequency - dependent complex dielectric function $\varepsilon(\omega)$ which describes the response of a 2D superconductor to an external time-dependent field. The measured $L_k(T)$ is renormalized from the BCS $L_{k0}(T)$ which is the $L_k(T)$ in the absence of the vortices: $L_{k0}(T)/L_k(T) = n_s/n_s^0 = \text{Re}[1/\varepsilon(\omega)]$. It is easy to show from[20,22] the following relation in high frequency limit:

$$\frac{L_k^{-1}(T)}{w\,\text{Re}\,s} = \frac{p(Y-1)}{2Y \ln Y} \qquad (6)$$

where $Y = (l_{\omega}/\xi_+)^2$. Both real and imaginary part of the $1/\varepsilon(\omega)$ are directly related to $Y$[20]. By solving Eq.6 for $Y(T)$ using $L_k^{-1}(T)$ and $\text{Re}\sigma(T)$ data, we obtained $Y(T)$ for different S1 samples and plotted as Y versus $T^{\omega}_{BKT}/T$ in Fig. 6.

From the Y(T) data, we thus found that the vortex diffusion constant D(T) is not linear with T at low temperature range as in the case for free vortices[20], but rather can be fitted with activation dependence $D(T) = D_0\exp[E_0/k_B(T_{c0}^{-1}-T^{-1})]$ due to pinning of vortex core[23]. The pinning energy, $E_0/k_B$=48 K, 1270 K and 3290 K for the 1, 2 and 3-UC S1 samples, can then be obtained. Notice, we found stronger thickness dependence of the pinning energy than $E_0/k_B(K)=260d(nm)$ in the BSCCO trilayers[17] and $E_0/k_B(K)=450d(nm)$ in the YBCO/PrBCO superlattices[37], probably due to different $\lambda(0)$ values of different samples as observed from the infrared measurements[38].

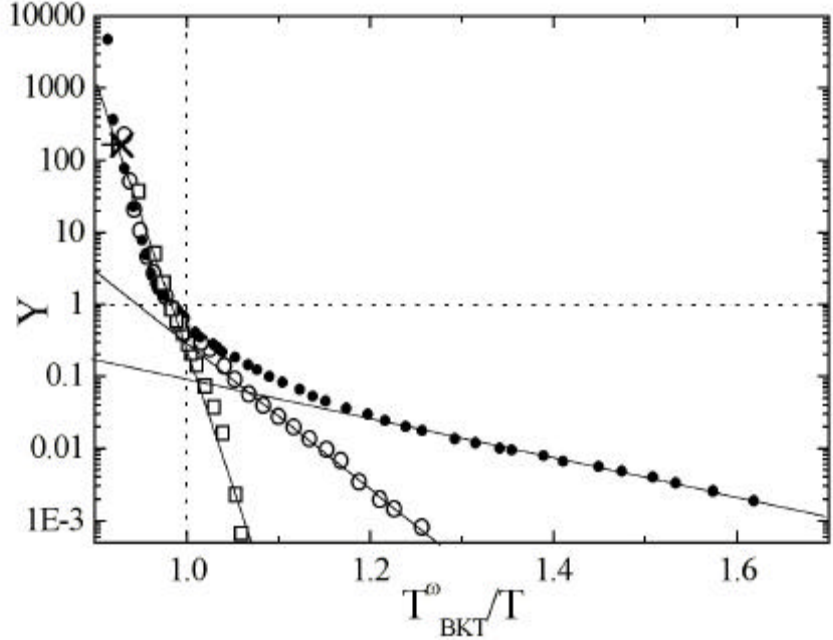

Fig.6. The Y(T) versus $T^{\omega}_{BKT}/T$ for 1- (closed circles), 2- (open circles) and 3- UC (open squares) S1 samples at 8 MHz. Big straight and tilted crosses shows the 500 MHz data for 2- and 3-UC S2 films, respectively.

We determined the $T^{\omega}_{BKT}$ as the peak position of $\sigma_1(T)$ point, which almost coincides with the point where Y=1. We also used the $T^{dc}_{BKT}$ as the point where $L_k^{-1}$(T) deviates from the square fit (see Fig. 3), as was used by Hebard *et al.*[39]. The ratio $T^{dc}_{BKT}/L_k^{-1}(T^{dc}_{BKT})$ thus equal to 25, 25, and 17 nHK for 1-, 2- and 3-UC S1 films, respectively, is about constant, which however is larger than the theoretical estimation $T^{dc}_{BKT}\lambda^2(T^{dc}_{BKT})/d = \phi_0^2/32\pi^2 k_B = 0.98$cmK (or $T^{dc}_{BKT}/L_k^{-1}(T^{dc}_{BKT}) = 12.3$ nHK)[22] ignoring small dynamic theory corrections of Ambegaokar *et al.*[21] to the renormalized coupling constant $K_R$. When $T > T^{dc}_{BKT}$, the $Y(T^{\omega}_{BKT}/T)$ deviates from exponential lines due to temperature dependence of $\xi_+(T)$ but collapse into one curve, indicating that the temperature dependence of $\xi_+(T^{\omega}_{BKT}/T)$ is the same for all samples. The 500 MHz data for $T^{\omega}_{BKT}$ for 2-UC and 3-UC S2 samples were converted to RF by multiplying Y=1 on the MW/RF frequency ratio since $Y\sim 1/\omega$. Notice very good agreement of these data with 8 MHz ones for S1 samples (see straight and tilted crosses in Fig. 6).

The Abrikosov vortex lattice parameter $a_v$ is the scale limiting the formation of vortex-antivortex pairs in magnetic field[40]. We can estimate the field $H_{ext}$ which destroys the vortex pair unbinding from the following relation: $l_0 \sim a_v = (\phi_0/H_{ext})^{1/2}$. This estimation gives $H_{ext} \approx 1.5$ mT at 10 MHz for 1UC and 2UC films in good agreement with the experimental results (see Fig. 3). Since $H_{ext}\sim\omega$, we expect that a larger field is required at higher frequencies in order to destroy the unbinding transition. In fact, the magnetic field which destroys the transition at RF, did not influence the transition at MW[24]. The effect was observed at higher field (above 5 mT) at MW, but apparently it is not due to the vortex-antivortex pair unbinding but rather the normal vortex formation in a magnetic field.

Rogers et al. argued that the vortex-pin interaction suppresses free vortex motion rather than vortex-antivortex interaction[17] for single unit cell BSCCO 2:2:1:2. However, the measurement were done at low frequencies where the pairs detected would have r > $\lambda_{eff}$. On the other hand, vortex pinning is also present in our samples because of the exponential dependence of D(T) observed. Nevertheless, the consistent picture of the frequency dependence of the data and the scaling behavior of the universal superfluid jump (Fig. 5) let us believe that vortex-antivortex interaction was indeed observed with the presence of pinning. Fisher pointed out that weak pinning by point defects does not affect the BKT transition[41], but it is not clear for the case of strong pinning.

## 4. CONCLUSIONS

In summary, we have compared our experimental results on ultrathin YBCO films with the extended dynamic theory for BKT transition and found that the vortex-antivortex pairs with short separation lengths are present. The unbinding of the vortex pairs were observable at high frequencies even though a true BKT transition is absent in the samples. Our results also indicate that part of the transition broadening in ultrathin YBCO films can be related to the dissociation of vortex-antivortex pairs.

## ACKNOWLEDGEMENTS

We are grateful to V.F. Gantmakher, A. Hebard, M. Chan, R. Huguenin, C. Lobb, P. Martinoli, D. van der Marel, D. Pavuna, C. Rogers, D.J. Scalapino, J.-M. Triscone, T. Venkatesan, X.X. Xi for stimulating discussions. This work was partially supported by Russian Scientific Programs: Superconductivity of Mesoscopic and Highly Correlated Systems (Volna 4G); Surface Atomic Structures (No.4.10.99) and Russian Ministry of Industry, Science and Technology (MSh-2169.2003.2), RFBR (No.02-02-16874-a), NSF DMR#0405502, CRDF (Award 3107) and by INTAS (No.01-0617).